\documentclass[unsortedaddress]{revtex4-1}

\usepackage{graphicx}
\usepackage{amssymb}
\usepackage{amsmath}
\usepackage{color} 

\newcommand{\red}[1] {{#1}}
\newcommand{\pd}[2]{\frac{\partial #1}{\partial #2}}

\usepackage[latin1]{inputenc}
\usepackage[english]{babel}
\usepackage{amsmath}
\usepackage{amssymb,amsfonts,textcomp}
\usepackage{array}
\usepackage{hhline}
\usepackage{graphicx}

\begin{document}

\title{Relaxation of a steep density gradient in a simple fluid: comparison between atomistic and continuum modeling.}

\author{Meisam Pourali}
\affiliation{Department of Physical Chemistry, School of Chemistry, College of Science,
University of Tehran, Tehran, Iran}

\author{Simone Meloni}\thanks{To whom correspondence should be addressed: simone.meloni@epfl.ch}
\affiliation{Laboratory of Computational Chemistry and Biochemistry, \'Ecole Polytechnique F\'ed\'erale de Lausanne, CH-1015 Lausanne, Switzerland}

\author{Francesco Magaletti}
\affiliation{Dipartimento di Ingegneria Meccanica e Aerospaziale, Universit\`a  La Sapienza, 
Via Eudossiana 18, 00184 Rome, Italy}

\author{Ali Maghari}
\affiliation{Department of Physical Chemistry, School of Chemistry, College of Science,
University of Tehran, Tehran, Iran}

\author{Carlo Massimo Casciola}
\affiliation{Dipartimento di Ingegneria Meccanica e Aerospaziale, Universit\`a  La Sapienza, 
Via Eudossiana 18, 00184 Rome, Italy}

\author{Giovanni Ciccotti}
\affiliation{Dipartimento di Fisica and CNISM, Universit\`a  La Sapienza, P.le A. Moro 5, 00185 Rome, Italy}

\begin{abstract}
We compare dynamical nonequilibrium molecular dynamics and continuum simulations of the dynamics of relaxation of a fluid system characterized by a non uniform density profile. Results match quite well as long as the lengthscale of density nonuniformities are greater than the molecular scale ($\sim 10$ times the molecular size). In presence of molecular scale features some of the continuum fields (e.g. density and momentum) are in good agreement with atomistic counterparts, but are smoother. On the contrary, other fields, such at the temperature field,  present very large difference with respect to reference (atomistic) ones. This is due to the limited accuracy of some of the empirical relations used in continuum models, the equation of state of the fluid in the example considered.

\end{abstract}

\maketitle

\section{Introduction}
Hydrodynamical phenomena are usually described in terms of fields, e.g. number/mass density,
$\rho\left (\textbf{x},t\right )$, momentum density, \textcolor{red}{$\boldsymbol{\pi}\left (\textbf{x},t\right )$},  
 energy density, $e\left (\textbf{x},t\right )$,  temperature, $T\left (\textbf{x},t\right )$, etc., where $\textbf{x}$
 is a point in the ordinary $\Re^3$ space.~\cite{hydroBook1,hydroBook2} $\rho\left (\textbf{x},t\right )$, 
$e\left (\textbf{x},t\right )$, and $\red{\boldsymbol{\pi}}\left (\textbf{x},t\right )$ obey 
conservation laws (see Sec.~\ref{sec:continuum_hydrodynamics}). This set of equations is not closed and, thus, they cannot be 
solved to obtain these fields. This problem is solved by supplementing the conservation laws with 
a set of empirical constitutive laws (e.g. Fick's law, Fourier's Law), 
which depend on the materials forming the system \emph{via} the value of certain transport coefficients. \red{In addition, it must be  assumed that the system is locally at the equilibrium}.
By solving the complete set of equations with given initial conditions one obtains the relevant fields at any
time $t$, thus fully characterizing the dynamics of the (continuum) system. 

This macroscopic description is obviously valid only when atomistic/molecular effects are 
not relevant, i.e. when the characteristic length and time scales of the process 
under investigation are \red{much} larger than the atomistic/molecular ones. However, also 
when these conditions are met, continuum hydrodynamics presents three problems. \red{First, it cannot be checked that the local equilibrium hypothesis holds.} Second, 
constitutive laws have a phenomenological origin, and their accuracy must be validated case by case.
Third, these equations requires to set the value of the associated transport coefficients in 
the given conditions (pressure, temperature, etc.), which might be not available. 

Alternatively, it is possible to use an \emph{ab initio} 
atomistic/molecular approach. In this case, the system is described in terms of 
its constituents, atoms and/or molecules, and time-dependent statistical mechanics is used to 
obtain the time evolution of the system. The relation between the macroscopic 
and microscopic description of hydrodynamics has been established more than 60 
years ago by Irving and Kirkwood~\cite{IrvingKirkwood}. The main ingredient of this formulation is the 
time-dependent Probability Density Function (PDF), $m\left (\Gamma, t\right )$  ($\Gamma$ 
is a point in the $\Re^{6N}$
 phase space, with $N$ number of particles in the system). Macroscopic fields 
 can be obtained as an ensemble average of suitable microscopic observables: $o\left (\textbf{x}, t\right ) = 
 \int\,d\Gamma \tilde o \left (\Gamma, \textbf{x}\right ) m\left (\Gamma, t\right )$, in which we assumed that the microscopic observable does not depend explicitly on time. Microscopic observables, 
have \red{ typically, the form} $\tilde o\left (\Gamma, \textbf{x}\right ) = \sum_{i = 1}^N \hat o_i\left (\Gamma\right ) 
 \delta\left (\textbf{r}_i - \textbf{x}\right )$, where $\hat o_i\left (\Gamma\right ) $ is the observable 
 related to $o\left (\Gamma, \textbf{x}\right )$ associated to the particle $i$. For example, 
 the macroscopic (mass) density field, $\red{\rho}\left (\textbf{x},t\right )$, is the ensemble average 
 over the time dependent PDF of the microscopic observable $\tilde \rho\left (\Gamma, \textbf{x}\right ) = 
 \sum_i \mu_i \delta\left (\textbf{r}_i - \textbf{x}\right )$, with $\mu_i$ mass of 
 the $i$-th particle and $\textbf{r}_i$ its position. \red{Analogously, $\boldsymbol{\pi}(\textbf{x},t)$ is the ensemble average of $\tilde{\boldsymbol{\pi}}\left (\Gamma, \textbf{x}\right ) =   \sum_i \textbf{p}_i \delta\left (\textbf{r}_i - \textbf{x}\right )$ where $\textbf{p}_i$ is the momentum of the i-th particle; and $e(\textbf{x},t)$ is the ensemble average of   ${\tilde e}\left (\Gamma, \textbf{x}\right ) =  \sum_i \left[\textbf{p}_i \cdot \textbf{p}_i/(2\mu_i) + 1/2\sum\limits_{i > j}\Phi \left(\textbf{r}_i,\textbf{r}_j \right) \right]\delta\left (\textbf{r}_i - \textbf{x}\right )$ where, for notational simplicity,  the particles are assumed to interact via a pair potential $\Phi \left(\textbf{r}_i,\textbf{r}_j \right)$.
 }

The advantage of the microscopic approach is that it relies on more fundamental principles and hypotheses, 
namely that the particles obey a suitable dynamics (Newton dynamics, Langevin
dynamics, etc.) driven by a suitable potential, and that the latter can be modeled by an empirical or 
\emph{ab initio} force field. 
No phenomenological laws, and associated transport coefficients, are needed. 
However, the microscopic approach presents other problems. A first, technical one, 
is that this approach is computationally expensive. In fact, the number of particles 
necessary to describe a given piece of matter is much larger then the number of 
grid points or finite elements used in the continuum methods. The second, more 
fundamental one, is the problem of determining the time-dependent PDF central to 
the Irving-Kirkwood microscopic formulation of hydrodynamics.

The interest for hydrodynamics at the nanoscale (i.e. the interest on phenomena occurring at smaller space and time scales)
on the one hand, and the exponential growth of the computational power, on the other hand,
make the atomistic approach more appealing. The shrinking of the space scale has a 
two-fold effect: first, the validity of the continuum description at this scale, in particular of the usual constitutive laws, is questionable. Moreover, in case they are still valid, the 
associated transport coefficients need to be determined. Second, the reduced size of these 
systems makes the problem treatable at the atomistic/molecular level.

From the atomistic standpoint, a still open question is how to compute, or sample, the relevant
time dependent PDF or, 
equivalently, estimate time-dependent macroscopic fields \emph{via} ensemble averages over 
it. Recently, some of the authors of the present work have proposed an approach 
to compute these ensemble average when $m\left (\Gamma, t\right )$ is the time evolution of an initial (conditional)
PDF having the form $w\left (\Gamma\right ) \prod_{\alpha=1}^{N_o} \delta\left (\tilde o_\alpha\left (\Gamma, \textbf{x}\right ) - o_\alpha\left (\textbf{x}\right )\right ) / \bar{ \mathcal{Q}}$, where $w\left (\Gamma\right )$ is the microscopic equilibrium PDF 
(e.g. the Boltzmann distribution $w\left (\Gamma\right ) = \exp[-\beta H\left (\Gamma\right )] / \int\,d\Gamma \exp[-\beta H\left (\Gamma\right )]$, $\bar{ \mathcal{Q}} = \int d\Gamma  w\left (\Gamma\right ) \prod_{\alpha=1}^{N_o} \delta\left (\tilde o_\alpha\left (\Gamma, \textbf{x}\right ) - o_\alpha\left (\textbf{x}\right )\right )$, with $H\left (\cdot\right )$ denoting the system Hamiltonian),  $\{o_\alpha\left (\textbf{x}\right )\}_{\alpha=1,N_o}$ is a set of (field) observables defining the initial 
macroscopic conditions. This 
method is based on the combination of the dynamical nonequilibrium molecular 
dynamics (D-NEMD) approach~\cite{D-NEMD,D-NEMD1,D-NEMD2}  with restrained MD (RMD)~\cite{TAMD,TAMC,RE-RMD}.  
RMD/D-NEMD has been already applied to the simple case of the relaxation of 
the interface between immiscible liquids~\cite{DNEMD-PCCP,DNEMD-PROCEEDINGS}, showing that, at a variance with ``standard'' 
nonequilibrium approaches (e.g.~\cite{Puhl:PhysicalReviewA:1989}), it produces results satisfying some fundamental symmetry properties of the system under investigation. 

The objective of this article is to compare the continuum, Navier-Stokes description of a simple, but non trivial, nanoscale hydrodynamic phenomenon with its ``exact'' RMD/D-NEMD counterpart.
In particular, we consider a single component system 
with an initial steep density gradient. The atomistic system is made of 
Lennard-Jones (LJ) particles, and its initial number density along the $x_1$ direction goes 
from $0.4$ to $0.8$ over $\sim 10$ particle radii (LJ units - see note~\footnote{LJ units consist in expressing lengths in $\sigma$, the particle radius in the LJ (see Sec.~\ref{sec:theory}), 
energies in $\epsilon$, the depth of the LJ potential well, and masses in  atomic
mass units. Thus, densities are expressed in $\sigma^{-3}$.} - are used throughout the text).

We consider this work a first step toward our 
attempt to establish an atomistic/continuum multiscale approach based on the 
RMD/D-NEMD formulation of microscopic hydrodynamics.

The manuscript is organized as follows. In Sec.~\ref{sec:theory}, we summarize 
the atomistic and continuum methods used in this article. In Sec.~\ref{sec:computational_setup} we describe the setup of our simulations. In Sec.~\ref{sec:results_discussion} 
we present results of atomistic and continuum simulations, and discuss their differences. Finally, in 
Sec.~\ref{sec:conclusions} we draw conclusions.

\section{Theoretical background}
\label{sec:theory}

\subsection{``\emph{Ab initio}'' hydrodynamics}
\label{sec:abinitio_hydrodynamics}
In classical mechanics the  state of a system is represented by a point, $\Gamma$, in the $6N$ phase space, consisting in the positions, $\textbf{r}$, and momenta, $\textbf{p}$, of all the particles. The probability density to be at the point $\Gamma$ at time $t$, $m\left (\Gamma, t\right )$, obey to the Liouville equation:
\begin{equation}
\label{eq:Liouville}
\frac{\partial m\left (\Gamma, t\right )}{\partial t} = - i L\left (\Gamma, t\right ) m\left (\Gamma, t\right )
\end{equation}

\noindent $i L\left (\Gamma, t\right )$ denotes the Liouville operator, whose action on $m\left (\Gamma, t\right )$ is $i L\left (\Gamma, t\right ) m\left (\Gamma, t\right ) = \{H\left (\Gamma, t\right ), m\left (\Gamma, t\right )\}$, with $H\left (\Gamma, t\right )$ the system Hamiltonian 
and $\{\cdot, \cdot\}$ Poisson brackets. Once initial conditions are given, Eq.~\ref{eq:Liouville} has a unique solution, which we denote by $U^\dag\left (t\right ) m\left (\Gamma, 0\right )$, with $U^\dag\left (t\right )$ time propagator  of the PDF and $m\left (\Gamma, 0\right )$ PDF of an initial ensemble.

The value of a (field) observable at time $t$ is then
\begin{eqnarray} 
\label{eq:macroscopic_field}
o\left (\textbf{x}, t\right ) &=&  \int\,d\Gamma\,\, \tilde o \left (\Gamma, \textbf{x}\right ) U^\dag\left (t\right ) m\left (\Gamma, t\right ) \nonumber \\
                      &=& \int\,d\Gamma \,\, U\left (t\right ) \tilde o \left (\Gamma, \textbf{x}\right )  m\left (\Gamma, 0\right )
\end{eqnarray} 

\noindent where we used the fact that  $U^\dag\left (t\right )$ is the adjoint of the time evolution operator,   $U\left (t\right )$, of particles' dynamics ($\Gamma\left (t; \Gamma \left (0\right ) \right) = U\left (t\right ) \Gamma\left (0\right )$).
The notation $U\left (t\right ) \tilde o \left (\Gamma, \textbf{x}\right )$ means that the observable $\tilde o \left (\cdot, \textbf{x}\right )$ is computed at the point in phase-space corresponding to the evolution at time $t$ of $\Gamma$, $\Gamma\left (t; \Gamma \right)$ (see note~\footnote{This can be proven expressing $\tilde o \left (\Gamma, \textbf{x}\right )$ as a Taylor series of $\Gamma$, and then applying to the so expressed observable the time evolution operator $U\left(t\right )$.}).  

Atomistic/molecular simulations can be used to estimate the ensemble average in the second row of Eq.~\ref{eq:macroscopic_field}.
If $m\left (\Gamma, 0\right )$ can be sampled by Monte Carlo (MC) or MD,  $o\left (\textbf{x}, t\right )$ can be obtained from the following estimator:
\begin{equation}
\label{eq:estimator}
   o\left (\textbf{x}, t\right ) = \lim_{M\rightarrow \infty} {1 \over M} \sum_{\nu=1}^M \tilde o \left (\Gamma\left (t; \Gamma_\nu\right ), \textbf{x}\right )
\end{equation}
Consistently with the notation introduced above, $\Gamma\left (t; \Gamma_\nu\right )$ is the time-evolution of $\Gamma_\nu$ at time $t$, and $\left\{\Gamma_\nu\right\}_{\nu=1,M}$ is a set of phase space points sampled from $m\left (\Gamma, 0\right )$. Thus, $\tilde o \left (\Gamma\left (t, \Gamma_\nu\right ), \textbf{x}\right )$ can be computed by standard MD started at $\Gamma_\nu$.

We will now focus on how to sample $m\left (\Gamma, 0\right )$. In some cases  this can be achieved by straightforward MC/MD. This is when, for example, the system is initially in equilibrium and some perturbation is turned on at time $t=0$~\cite{D-NEMD,D-NEMD1,D-NEMD2,thermalNEMD,otherNEMD}. Another example is when the system is initially in a stationary non-equilibrium condition, for example when there is a temperature gradient, and then either a perturbation is turned on or the source of the stationary non-equilibrium is turned off~\cite{MauroLorenzo}. 
However, when studying hydrodynamic phenomena, often the initial condition is the conditional PDF of a set of macroscopic fields, $\left \{\tilde o_\alpha\left (\Gamma, \textbf{x}\right ) \right \}_{\alpha=1,n}$:

\begin{equation}
 m\left (\Gamma, 0\right ) = m\left (\Gamma; \prod_{\alpha=1}^{N_o} \delta\left (\tilde o_\alpha\left (\Gamma, \textbf{x}\right ) - o_\alpha \left (\textbf{x}\right )\right ) \right )
\end{equation}

\noindent frequent is the case in which the system is also attached to a thermostat. In this case

\begin{equation}
\label{eq:conditionalPDF}
m\left (\Gamma; \prod_{\alpha=1}^{N_o} \delta\left (\tilde o_\alpha\left (\Gamma, \textbf{x}\right ) - 
o_\alpha\left (\textbf{x}\right )\right ) \right ) = {\exp[-\beta H\left (\Gamma\right )] \prod_{\alpha=1}^{N_o} \delta\left (\tilde o_\alpha\left (\Gamma, \textbf{x}\right ) - o_\alpha\left (\textbf{x}\right )\right ) \over \int d\Gamma \exp[-\beta H\left (\Gamma\right )] \prod_{\alpha=1}^{N_o} \delta\left (\tilde o_\alpha\left (\Gamma, \textbf{x}\right ) - o_\alpha\left (\textbf{x}\right )\right )} 
\end{equation}

\noindent  This conditional PDF can be sampled by RMD~\cite{TAMD,TAMC,DNEMD-PCCP,DNEMD-PROCEEDINGS,RMD-JSTATPHYS}. 

In RMD atoms are driven by a potential $\tilde V\left (\textbf{r}, \{o_\alpha\left (\textbf{x}\right )\}_\alpha\right )$, function of  $\textbf{r}$ and functional of $\{o_\alpha\left (\textbf{x}\right )\}_\alpha$.  $\tilde V\left (\textbf{r}, \{o_\alpha\left (\textbf{x}\right )\}_\alpha\right )$
consists of the sum of the physical potential, $V\left (\textbf{r}\right )$, and the restraining potential $\int d\textbf{x} \sum_\alpha (k_\alpha/2) \,\, \left (\tilde o_\alpha\left (\Gamma, \textbf{x}\right ) - o_\alpha\left (\textbf{x}\right )\right )^2$. To make this scheme practical, the ordinary $\textbf{x}$ space is discretized. Thus, the confining potential reads\\ $\sum_\alpha \sum_l (k_\alpha/2)\,\, \left (\tilde o_\alpha\left (\Gamma, \textbf{x}_l\right ) - o_\alpha\left (\textbf{x}_l\right )\right )^2$, where $l$  runs over the $N_M$ grid points of the mesh used to discretize the $\textbf{x}$-space. An NVT MD driven by this potential samples the PDF \\$w_{\tilde V}\left (\Gamma\right ) =$  $\exp[-\beta H\left (\Gamma\right )]\,\, \prod_\alpha \exp[-\beta \sum_l (k_\alpha/2) \left (\tilde o_\alpha\left (\Gamma, \textbf{x}_l\right ) - o_\alpha\left (\textbf{x}_l\right )\right )^2] /$ $ \tilde{ \mathcal{Q}}$, where the subscript $\tilde V$ of the PDF indicates that it is relative to the restrained ensemble, and  $\tilde {\mathcal{Q}}$ = $\int d\Gamma\,\,\exp[-\beta H\left (\Gamma\right )] \,\,\prod_\alpha \exp[-\beta \sum_l (k_\alpha/2) \left (\tilde o_\alpha\left (\Gamma, \textbf{x}_l\right ) - o_\alpha\left (\textbf{x}_l\right )\right )^2]$ is the associated partition function. Eq.~\ref{eq:conditionalPDF} is recovered in the limit $\beta k_\alpha \rightarrow \infty$, in which \\ $w_{\tilde V}\left (\Gamma\right ) \rightarrow m\left (\Gamma; \prod_{\alpha=1}^{N_o} \delta\left (\tilde o_\alpha\left (\Gamma, \textbf{x}\right ) - o_\alpha\left (\textbf{x}\right )\right ) \right )$~\cite{TAMD,TAMC,DNEMD-PCCP,DNEMD-PROCEEDINGS,RMD-JSTATPHYS}, proving that RMD samples the desired conditional PDF.

\subsection{Continuum hydrodynamics}
\label{sec:continuum_hydrodynamics}
Continuum fluid dynamics concerns the evolution of the basic (macroscopic) fields $\rho(\textbf{x},t)$, $\boldsymbol{\pi}(\textbf{x},t)$, $e(\textbf{x},t)$. 
They obey conservation equations of the form
\begin{equation}\label{eq.mass}
\pd{\rho}{t} + \boldsymbol{\nabla} \cdot \boldsymbol{\pi} = 0 \,,
\end{equation}
\begin{equation}\label{eq.momentum}
\pd{\boldsymbol{\pi}}{t} + \boldsymbol{\nabla} \cdot \left(\boldsymbol{\pi}\otimes\textbf{v}\right) = \boldsymbol{\nabla} \cdot \boldsymbol{\tau} \,,
\end{equation}
\begin{equation}\label{eq.energy}
\pd{e}{t} + \boldsymbol{\nabla} \cdot \left(\textbf{v} e\right) = \boldsymbol{\nabla} \cdot \left(\boldsymbol{\tau}\cdot \textbf{v} - \textbf{q}\right) \,,
\end{equation}
\noindent where we omitted the dependence on $\textbf{x}$ and $t$ to make the notation shorter. In Eqs.~\ref{eq.momentum}-\ref{eq.energy}  $\textbf{v}(\textbf{x},t) = \boldsymbol{\pi}(\textbf{x},t)/\rho(\textbf{x},t)$, $\boldsymbol{\tau}(\textbf{x},t)$, and $\textbf{q}(\textbf{x},t)$ are the velocity, stress tensor and energy flux fields, respectively. It is worth stressing that the above equations can be derived from a microscopic description of  the underlying atomistic system, starting from the Liouville equation (Eq.~(\ref{eq:Liouville})). This makes a direct connection between continuum and atomistic approaches. There is however a crucial difference.
In the atomistic description the basic unknown is the PDF $m(\Gamma,t)$. In this case 
the Irwing-Kirkwood procedure provides a microscopic expression for stress tensor and energy flux in terms of  
$m(\Gamma,t)$. 
In continuum mechanics, instead, $m(\Gamma,t)$ is not accessible and, since the three Eqs.~\ref{eq.mass}-\ref{eq.energy}  involve five unknown fields, the system is not closed.
This difficulty is circumvented by adding suitable phenomenological constitutive relations which describe the rheology of the material, \red{together with the hypothesis of local equilibrium}. 
In specifying the constitutive relations, certain general constraints must be satisfied, that we briefly review for the reader's convenience, see e.g. \cite{de2013non} for additional details.

Here we will assume a homogeneous and isotropic fluid
governed by linear constitutive laws. 
Memory effects, such as the dependence of the stress on the past deformation history of the material, will be excluded.
Thus, constitutive relations will depend only on the present state of the system as identified by the basic fields.
In principle constitutive relations may depend non-locally on the basic fields like, e.g., when the energy flux is associated to radiation phenomena.  Even such non-local behavior will be taken out of consideration here, by requiring that the constitutive relations express the auxiliary fields in terms of the {\sl almost} local behavior of the basic fields, 
i.e. assuming the dependence of stress and energy flux on the basic fields and their gradients.

Further,  being interested in fluids,  no dependence on deformation is allowed and we shall assume the co-variance of the model with respect to rigid changes of reference frame. This rules out dependence on velocity as such and on the antisymmetric part of the
velocity gradient $\boldsymbol{\Omega} = 1/2\left(\boldsymbol{\nabla} \textbf{v} - \boldsymbol{\nabla} \textbf{v}^T \right)$ which can always be made to locally vanish by a suitable angular velocity
of the reference frame. In this context, the so-called Newtonian fluids obey linear relations $\boldsymbol{\Sigma} = \boldsymbol{\tau} + p \textbf{I}\propto \textbf{E}$,  with $p$ the thermodynamic pressure and $\textbf{I}$ the identity,  where the viscous component of the stress, $\boldsymbol{\Sigma}$, depends linearly on the velocity deformation rate $\textbf{E} = 1/2\left(\boldsymbol{\nabla} \textbf{v} + \boldsymbol{\nabla} \textbf{v}^T \right)$, and $\textbf{q} \propto \boldsymbol{\nabla} T$, being $T(\textbf{x},t)$ the temperature field to be commented  on in a while.

Despite of the above assumptions, we are still left with a substantial freedom in choosing specific (linear) constitutive relations.
However basic limitations imposed by thermodynamics need \red{to be} satisfied, namely that the entropy variation associated with \emph{any} macroscopic part of the system should be larger than the entropy flux \red{entering the domain} (Gibbs-Duhem inequality),
\begin{equation}
{\dot S} = \frac{d}{dt}\int_{\cal D} s\,\, d\textbf{x} =  \int_{\cal D} \frac{\partial s}{\partial t}\,\, d\textbf{x} \ge \red{-} \int_{\partial \cal D}  \boldsymbol{\Phi_s} \cdot \textbf{n}\,\, dS,
\end{equation}
where $s$ (shorthand for $s\left(\textbf{x},t\right)$) is the entropy field, $\textbf{n}$ is the outward normal to the boundary $\partial \cal D$ enclosing the considered portion $\cal D$ of the flow domain, and  $\boldsymbol{\Phi_s} = \textbf{v} \, s + \textbf{q}/T$ is the entropy flux.
An equivalent form, that we will use below, is

\red{
\begin{equation}
\label{eq:GibbsDuhem}
\int_{\cal D} \left( \frac{\partial s}{\partial t} + \boldsymbol{\nabla} \cdot \boldsymbol{\Phi_s}\right) d\textbf{x} \ge 0 \, ,
\end{equation}
\noindent The next step, then, is deriving an evolution law for the entropy density field, $\partial s\left(\textbf{x},t\right)/\partial t$. The starting point for this derivation are the conservation laws, Eqs.~(\ref{eq.momentum}) and (\ref{eq.energy}).}
The energy density (per unit volume) can be expressed as the sum of two terms: $e = K_m + u$, where $u$ is \red{identified with the thermodynamic potential ``internal energy density'',} and $K_m = 1/2 \rho |\textbf{v}^2| = |\boldsymbol{\pi}|^2/(2 \rho)$ \red{is the} macroscopic kinetic energy density. 
\red{
The evolution equation for the macroscopic kinetic energy, ${\partial K_m}{/\partial t} + \boldsymbol{\nabla} \cdot \left(\textbf{v} K_m \right) = \textbf{v} \cdot \left(\boldsymbol{\nabla} \cdot \boldsymbol{\tau} \right)$,  follows from momentum conservation, Eq.~(\ref{eq.momentum}), by scalar multiplying by 
$\textbf{v}$. The evolution equation for the  internal energy density $u$ is then obtained  by subtracting the equation for $K_m$ from the equation for the total energy density, Eq.~(\ref{eq.energy}),
}
\begin{equation}\label{eq.internal_energy}
\pd{u}{t} + \boldsymbol{\nabla} \cdot \left(\textbf{v} u\right) = \boldsymbol{\nabla} \textbf{v} : \boldsymbol{\tau}   -  \boldsymbol{\nabla} \cdot \textbf{q} \ .
\end{equation}
In \red{thermodynamic equilibrium} the specific 
internal energy, ${\hat u}= u/\rho$, is a function of \red{the} mass density and of the specific entropy $\hat s$ that enters the picture through the 
relation ${\hat u} = {\hat u}(\rho,{\hat s})$. The classical way to extend the \red{thermodynamic equilibrium} to  (slightly) nonequilibrium conditions is by postulating that\red{, locally,} the fundamental thermodynamic relation holds in terms of local values of the fields,
${\hat u}(\textbf{x},t) = {\hat u}\left[\rho(\textbf{x},t),{\hat s}(\textbf{x},t)\right]$. This extension leads to the definition of the temperature field $T(\textbf{x},t) = \partial {\hat u}/\partial {\hat s}\vert_\rho(\textbf{x},t)$ and of the pressure field $p(\textbf{x},t) = -\partial  {\hat u}/\partial v \vert_{\hat s}(\textbf{x},t) =
\rho^2 \partial  {\hat u}/\partial \rho \vert_{\hat s}(\textbf{x},t)$, where $v = 1/\rho$ is the specific volume. 
To make the successive manipulation easier,  Eq.~(\ref{eq.internal_energy}) is conveniently recast in terms of $\hat u$ as $\rho D{\hat u}/Dt = \boldsymbol{\nabla} \textbf{v} : \boldsymbol{\tau}   -  \boldsymbol{\nabla} \cdot \textbf{q}$, where we made use of mass conservation, Eq.~(\ref{eq.mass}), and the symbol
$D/Dt = \partial /\partial t + \textbf{v} \cdot \boldsymbol{\nabla}$ is commonly called the material derivative.

Substituting the fundamental thermodynamic relation \red{${\hat u}(\rho,{\hat s})$} in the above equation, considering that 
$D{\hat u}/Dt = \partial {\hat u}/\partial \rho\vert_{{\hat s}} D\rho/Dt + \partial {\hat u}/\partial {\hat s}\vert_{\rho} D{\hat s}/Dt = p/\rho^2 D\rho/Dt + T D{\hat s}/Dt$, one  gets the evolution equation for the specific entropy,
\begin{equation}
\rho T \frac{D {\hat s}}{Dt} =  \left(\boldsymbol{\tau}+p\textbf{I}\right):\boldsymbol{\nabla}\textbf{v} - \boldsymbol{\nabla} \cdot \textbf{q} \ .
\end{equation}
Considering that $\boldsymbol{\nabla} \cdot \textbf{q}/T =  \boldsymbol{\nabla} \cdot \left(\textbf{q}/T\right) -   \textbf{q} \cdot \boldsymbol{\nabla}\left(1/T \right)$,
the equation for the specific entropy can be rewritten in terms of the entropy density $s = \rho {\hat s}$, 
\begin{equation}
\label{eq:sEvolution}
\pd{s}{t} + \boldsymbol{\nabla} \cdot \boldsymbol{\Phi_s} = \frac{\boldsymbol{\Sigma}:\boldsymbol{\nabla}\textbf{v}}{T} -\frac{\textbf{q}\cdot\boldsymbol{\nabla}T}{T^2}\, .
\end{equation}
%

Eq.~\ref{eq:sEvolution} is substituted into Eq.~\ref{eq:GibbsDuhem}
%
%
%
\red{
$$
\int_{\cal D} \left( \frac{\partial s}{\partial t} + \boldsymbol{\nabla} \cdot \boldsymbol{\Phi_s}\right) d\textbf{x} = 
\int_{\cal D} \left(  \frac{\boldsymbol{\Sigma}:\boldsymbol{\nabla}\textbf{v}}{T}  -\frac{\textbf{q}\cdot\boldsymbol{\nabla}T}{T^2}
\right) d\textbf{x} \ge 0 \, .
$$
Given the arbitrariness of the domain $\cal D$, }
\begin{equation}
\label{eq.Gibbs_Duhem}
 \frac{\boldsymbol{\Sigma}:\boldsymbol{\nabla}\textbf{v}}{T} -\frac{\textbf{q}\cdot\boldsymbol{\nabla}T}{T^2} \ge 0 
\end{equation}
\noindent should be valid everywhere over \red{the flow domain}, with equality holding when the system is at thermodynamic equilibrium. 
In classical books on continuum thermodynamics, the left hand side of the inequality is called \emph{entropy production}.
For an isotropic fluid, the so-called Curie principle 
(see note \footnote{
The constitutive equations express the \emph{thermodynamic fluxes}, in our case the heat flux $\textbf{q}$ and the viscous component of the stress tensor $\boldsymbol{\Sigma} = \boldsymbol{\tau} + p \boldsymbol{I}$, as a linear combination of the \emph{thermodynamic forces}, here $\boldsymbol{\nabla} T$ and $\boldsymbol{\nabla} \textbf{v}$. In principle, each component of the fluxes could depend on all components of the forces. However, in presence of spatial symmetries, the Curie principle enforces certain constraints. For an isotropic fluid, the invariance of the phenomenological equations to rotations~\cite{de2013non} implies that fluxes of a given nature (e.g. scalars, polar vectors, axial vectors or symmetric tensors) may only depend on forces of corresponding nature. In our case, the heat flux depends only on the temperature gradient while the viscous stress depends only on the symmetric part of the velocity gradient.  This symmetry property  entails the decomposition of the entropy source into two independent components that, separately, should obey the requirement of positive definiteness.
})  
shows that thermodynamic fluxes of a given tensorial order can only depend on thermodynamic forces of the same order, i.e.
the symmetric viscous stress tensor only depends on the symmetric part of the velocity gradient $\boldsymbol{E}$,
$\boldsymbol{\Sigma} = \boldsymbol{\Sigma}\left(\boldsymbol{E}\right)$, and the heat flux only depends on the temperature gradient, $\textbf{q} = \textbf{q}\left(\boldsymbol{\nabla} T \right)$.
The positive definiteness of the two terms ${\boldsymbol{\Sigma}:\boldsymbol{\nabla}\textbf{v}}/{T} $ and $-{\textbf{q}\cdot\boldsymbol{\nabla}T}/{T^2}$ guarantees that the condition of Eq.~\ref{eq.Gibbs_Duhem} is satisfied.  
\red{
Taking into account the symmetry of the stress tensor, the most general linear dependence of the viscous stress  $\boldsymbol{\Sigma}$ on the symmetric part of the velocity gradient is (see note \footnote{The most general linear expression relating viscous stress and symmetric part of the velocity gradient reads
$\boldsymbol{\Sigma} = \boldsymbol{A} \left(\boldsymbol{\nabla} \textbf{v} +  \boldsymbol{\nabla} \textbf{v}^T\right)$, where $\boldsymbol{A}$ is a fourth order tensor. Invariance to rotation reduces the form of the tensor to $A_{ijkl} = a_1 \delta_{ij} \delta_{kl} + a_2 \delta_{il} \delta_{kj} +
a_3 \delta_{ik} \delta_{jl}$. Contraction with the symmetric part of the velocity gradient leads to
$\Sigma_{ij}= \left(a_1 \delta_{ij} \delta_{kl} + a_2 \delta_{il} \delta_{kj} +
a_3 \delta_{ik} \delta_{jl}\right) \left(\partial v_k/\partial x_l + \partial v_l/\partial x_k\right) = \lambda \partial v_k/\partial x_k \delta_{ij} + \mu \left(\partial v_i/\partial x_j + \partial v_j/\partial x_i \right)$, where the first and second viscosity coefficients are $\mu = a_2+a_3 $ and $\lambda = 2 a_1$, respectively.
})
}
\begin{equation}
\label{eq:momentumFlux}
\boldsymbol{\Sigma} =
2 \mu \boldsymbol{E} + \lambda   \textbf{I}  \textrm{Tr}\left( \boldsymbol{E} \right) \,,
\end{equation}
where  $\textrm{Tr} \left( \boldsymbol{E} \right) = \boldsymbol{\nabla} \cdot \textbf{v}$. 
The first and second viscosity coefficients must satisfy  $\mu\ge 0$ and $\lambda \ge - 2/3 \mu$, respectively,  where the inequalities  follow from  $\boldsymbol{\Sigma}(\boldsymbol{E}):\boldsymbol{E}  \ge 0$~\red{(see note \footnote{
Using Eq.~(\ref{eq:momentumFlux}) and decomposing a symmetric tensor into the sum of its  traceless and spheric components, e.g. 
$\boldsymbol{\Sigma} = \boldsymbol{\Sigma}_0 + 1/3 \textrm{Tr}\left(\boldsymbol{\Sigma}\right) \boldsymbol{I}$,
the entropy source associated to the viscous stress can be rewritten as $\boldsymbol{\Sigma} : \boldsymbol{E} = \boldsymbol{\Sigma}_0:\boldsymbol{E}_0 + 1/3 \textrm{Tr}\left(\boldsymbol{\Sigma}\right) \boldsymbol{I}: 1/3 \textrm{Tr}\left(\boldsymbol{E}\right) \boldsymbol{I}$,
where $\textrm{Tr}\left(\boldsymbol{\Sigma}\right) = \left(2 \mu + 3\lambda\right) \boldsymbol{\nabla} \cdot \textbf{v}$ and $\boldsymbol{\Sigma}_0 = 2 \mu 
\boldsymbol{E}_0$. It follows $\boldsymbol{\Sigma} : \boldsymbol{E} = 2 \mu \boldsymbol{E}_0 : \boldsymbol{E}_0 + \left(\lambda + 2/3 \mu \right) \left(\boldsymbol{\nabla} \cdot \textbf{v}\right)^2 \ge 0$, which can be satisfied for any $\boldsymbol{E}$  when $\mu \ge 0$ and $\lambda + 2/3 \mu \ge0$.
}).
}
Concerning the energy flux, its general expression for an isotropic fluid corresponds to the classical Fourier \red{law}
\begin{equation}
\label{eq:energyFlux}
\textbf{q} = -k \boldsymbol{\nabla} T\, ,
\end{equation}
where the requirement that the flux of energy should be accompanied by entropy production, $-\textbf{q}(\boldsymbol{\nabla}T)\cdot\boldsymbol{\nabla}T\, \ge\, 0$, calls for a positive thermal \red{conductivity}, $k > 0$. 

Once the constitutive relation just described are introduced in the conservation equations for the basic fields, a closed system of equation results \red{(equations of motion, EoM)}, provided that suitable equations of state \red{(EoS)} are supplemented (typically EoS are used in the form of pressure and internal energy as a function of temperature and density, $p(\textbf{x},t) = p\left[\rho(\textbf{x},t), T(\textbf{x},t) \right]$, $u(\textbf{x},t) = u\left[\rho(\textbf{x},t), T(\textbf{x},t) \right]$).

Initial conditions on the basic fields, or equivalent information given, e.g., in terms of initial fields of density, velocity and temperature, needs to be prescribed to specify the initial macrostate. Boundary conditions along the whole boundary are also required for the momentum and the energy equation.
They are given prescribing velocity  (e.g. no-slip condition at solid boundaries) and temperature at the boundaries. Other kinds of boundary conditions are also common, such as assigning the contact force per unit surface area (traction) $\textbf{t} = \boldsymbol{\tau} \cdot \textbf{n}$ or the heat flux $\textbf{q} \cdot \textbf{n}$.
Mass conservation requires $\rho$ to be specified at inlet boundaries, where $\textbf{v} \cdot \textbf{n} < 0$ (the normal points outwards).

For the fluid constituted by a system of Lennard-Jones particles discussed below several empiric EoS obtained from atomistic simulations exist, e.g. the Johnson-Zollweg-Gubbins EoS~\cite{JohnsonZollwegGubbins}. Transport coefficients can be obtained from bulk atomistic simulations as well.

\section{Computational setup}
\label{sec:computational_setup}

\subsection{Atomistic simulations}
\label{sec:atomistic_setup}

Our system is a fluid of 20522 LJ particles in a $38\,\times\,30\,\times\,30$ triperiodic simulation box, corresponding to an average density of $\bar \rho \sim 0.6$. 

\begin{figure}[h!]
 \includegraphics[width=15cm]{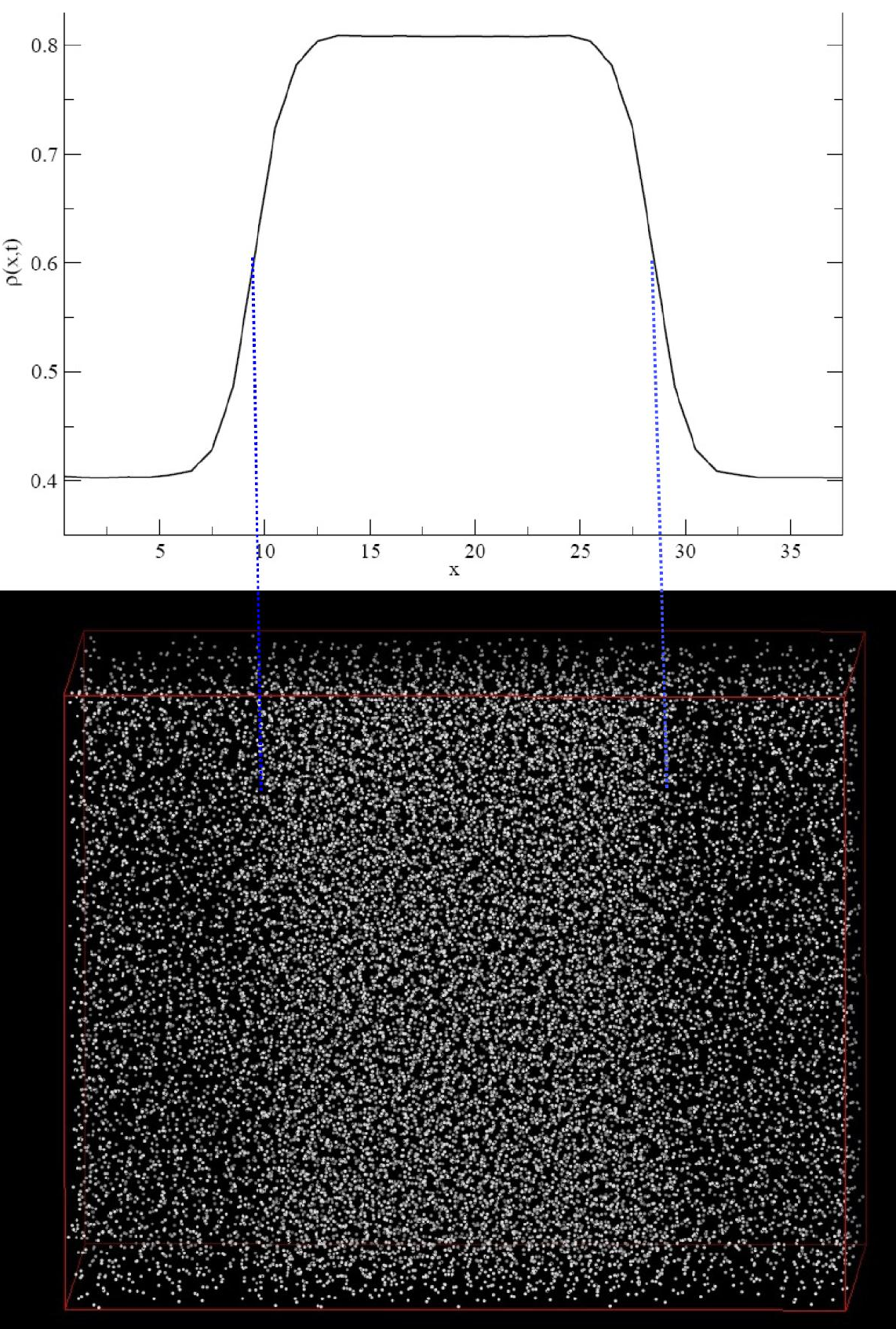}
 \caption{Top: initial density profile $\rho\left ( \textbf{x}, t\right )$. Bottom: one of the $600$ samples (atomic configuration) extracted from the initial conditional PDF.} 
 \label{fig:initial_rho}
\end{figure}

RMD simulations for sampling the initial conditional PDF are performed evolving the atoms according to the Nos\'e-Hoover chains~\cite{NHC} EoM at  temperature $T =1.5$. $T$  and $\bar \rho$, which lies in the supercritical domain of the LJ phase diagram, have been chosen to prevent possible phase transition. The initial macroscopic condition consists in a double (mirrored) $s$-shaped density profile along the $x_1$ direction (see Fig~\ref{fig:initial_rho}). More in detail:

\begin{equation}
\begin{array}{ll}
\rho(x_1, 0) = {1 \over 2} (\rho_1 + \rho_2) + {1 \over 2} (\rho_1 - \rho_2) \tanh[a(\,x_1-x_1^a)]; & x_1 \in  [0, 19) \\
\rho(x_1, 0) =  {1 \over 2} (\rho_1 + \rho_2) + {1 \over 2} (\rho_1 - \rho_2) \tanh[-a(\,x_1-x_1^b)]; & x_1 \in [19, 38) 
\end{array} 
\end{equation}

\noindent where $x_1^a = 9.5$ and $x_1^b = 28.5$ determine the positions of the two ``interfaces'', and $a = 2/\delta = 0.57$, with $\delta = 3.5$ the thickness of the interface. $\rho_1 \sim 0.8$  and $\rho_2 \sim 0.4$ are the initial values of density in the high and low density domains, respectively. There is no break of translational symmetry along $x_2$ and $x_3$.
The ordinary $\textbf{x}$-space is discretized by a $38 \,\times\,1\,\times\,1$ points mesh, i.e. it is discretized in slices along the $x_1$ direction. The single point discretization along $x_2$ and $x_3$ is consistent the symmetry of the system. Grid points of this discretization are denoted by the symbol $x_1^{(j)}$, with $j=1, 38$. The microscopic number density field on the grid points is defined as the average of the density field over the corresponding slices:  $\bar \rho\left (\Gamma, x_1^{(j)}\right ) =  1/\Omega^{(j)} \int_{\Omega^{(j)}} d\textbf{s}\,\, \sum_{i=1}^N \delta\left (\textbf{r}_i - \textbf{s}\right )$. This definition, however, is not suitable for RMD because it gives rise to impulsive forces coming from the restraining term when one particle moves from one slice to another. This problem is solved by resorting to an approximated (``mollified'') definition of the density, in which we replace the Dirac $\delta$-function with a Gaussian function:

\begin{eqnarray}
{\tilde \rho}_\epsilon (\textbf{r}, x^j_1) = {1 \over \Omega^{(j)}}\, \sum_{i = 1}^N \int_{\Omega_j} d\textbf{s}\,\,  g_\epsilon(|\textbf{r}_i - \textbf{s}|)
\end{eqnarray}

\noindent where $g_\epsilon(\cdot)$ is a gaussian function centered at $\textbf{r}_i$, position of the $i$th atom,  of variance $\epsilon$.  ${\bar \rho}(\textbf{r}, x^j_1) = \lim_{\epsilon \rightarrow 0} {\tilde \rho}_\epsilon(\textbf{r}, x^j_1)$. 
In practice, we set $\epsilon = 0.5$, which is small on the macroscopic scale but large enough to give a smooth atomistic force. 

\begin{figure}[h!]
 \includegraphics[width=15cm]{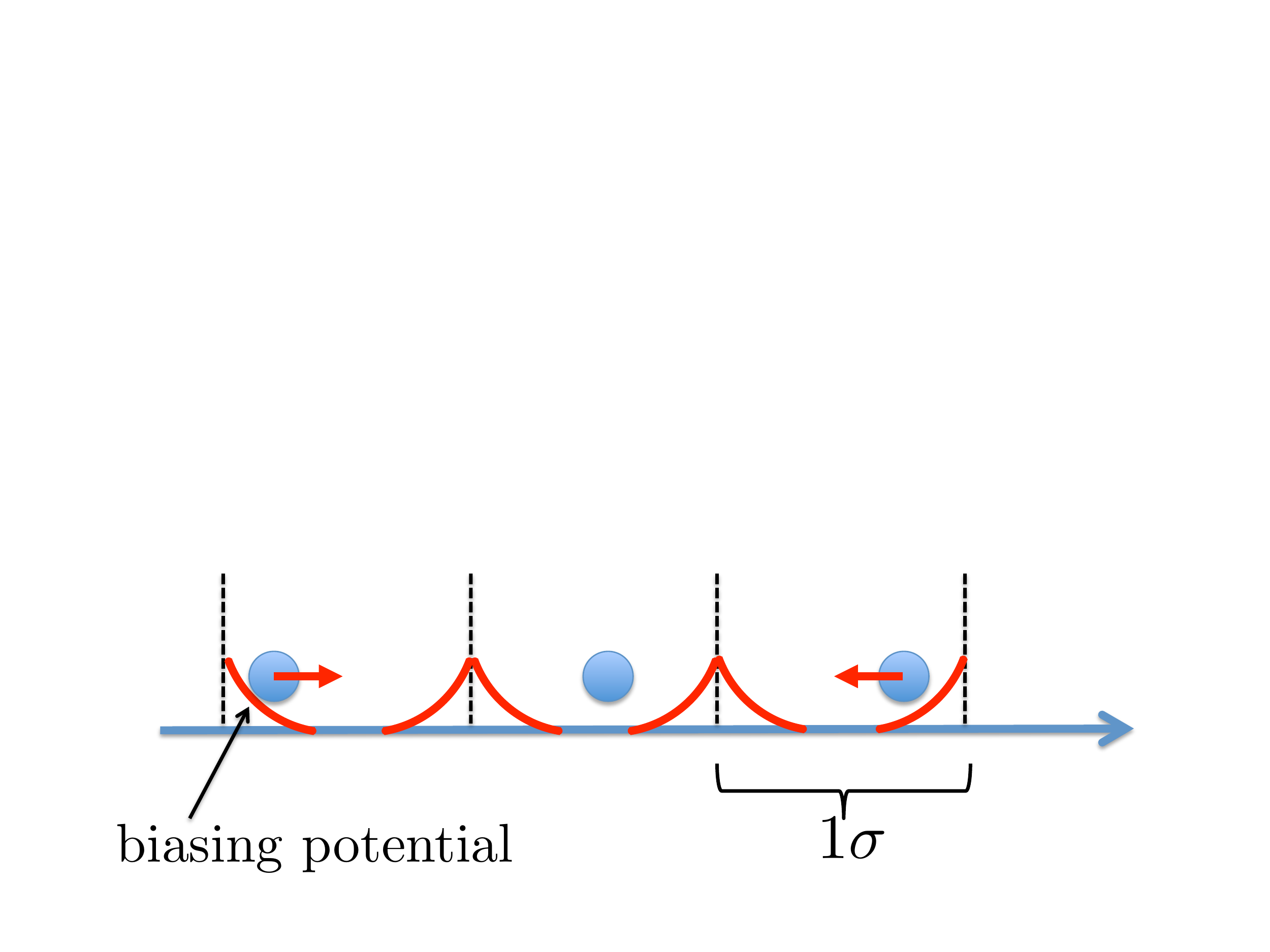}
 \caption{Sketch of the biasing potential (red lines) acting on the particles in the case of mollification of the atomistic density field. The vertical dashed lines denote the boundaries of slices discretizing the ordinary $\Re^3$ space. The red arrow denote the direction of the biasing force acting on the atoms.} 
 \label{fig:SoftWalls}
\end{figure}

The restraint potential, and the corresponding force, have the effect of preventing the variation of density in a slice out of its target value. Thus, for a slice with the associated density at the target value,  the restraint potential has the effect of keeping the particles within it (see note \footnote{In principle, the density of a slice can remain constant also if two particles cross its boundaries in opposite directions at the same time, one exiting and the other entering. However, this synchronous process is unlikely, and thus the effect of the restraint potential is confining particles within a slice, as described in the text.}).  In practice, particles in a slice are confined between soft walls (see Fig.~\ref{fig:SoftWalls}). We will discuss the effects of density mollification in the result section. 

To run RMD simulations we implemented the density field restraint in the PLUMED code~\cite{plumed}. PLUMED is a ``driver'' that allows to perform advanced sampling simulations, and must be used in combination with a ``standard'' MD code, LAMMPS~\cite{lammps} in the present case.

The sample of the initial conditional PDF consists of $600$ phase space points extrated from a $1200000$-steps long RMD simulation at  $T = 1.5$. These points are evolved in time by numerically integrating Newtonian (i.e. constant energy, volume and number of particles) EoM, thus obtaining the set $\{\Gamma\left (t, \Gamma_i\right)\}_i$. Then, using the estimator of Eq.~\ref{eq:estimator}, we compute any field $o\left (\textbf{x}, t\right )$ of interest.

\subsection{Continuum simulations}
\label{sec:continuum_setup} 
To solve the set of conservation equations (Eqs.~\ref{eq.mass}-\ref{eq.energy}) and constitutive laws (Eqs.~\ref{eq:momentumFlux} and \ref{eq:energyFlux}) we use a finite difference scheme specialized for compressible flows.
\red{The initial conditions for the continuum calculations are prescribed as  macroscopic fields obtained from atomistic simulations \emph{via} ensemble average of microscopic field-like observables, as explained in Sec.~\ref{sec:theory}. In the configuration we address, the basic fields are function of $x_1$ only, and are periodic along this direction. This implies that all fields will remain one-dimensional and periodic along the evolution. This allows us to solve 1D continuum equations with PBC consistent with atomistic simulations. }
 The spatial domain is discretized with a uniform grid of 190 cells of width $\Delta x = 0.2$ (though unusual,  LJ units will be used also in the continuum context). At every time  $t$, the relevant fields are computed at the centers of the cells
$f_i(t)=f(x_i,t)$ ($f$ \red{stands for anyone of the fields of interest}), with $x_i, \; i=0,1,...,190,$ (center cell locations). 
\red{The spatial second derivatives are computed according to }the second order ``central point'' approximation: 
 $\partial^2 f/\partial x_1^2 \vert_{x_i} = (f_{i+1}-2f_i+f_{i-1})/\Delta x^2 + {\cal{O}}\left(\Delta x^2\right)$. 
 \red{The treatment of the first derivatives is less straighforward.} Here we use the so-called Weighted Essentially Non-Oscillatory (WENO) procedure~\cite{shu_review}, which allows preventing spurious numerical oscillations associated to high order interpolation across discontinuities  (Gibbs phenomenon), \red{i.e. to achieve high order formal accuracy in smooth regions while maintaining stable, non-oscillatory and sharp discontinuity transitions.}

\red{After spatial discretization, continuum EoM reduces to a system of Ordinary Differential Equations (ODEs - three equations per cell). These are numerically integrated with a suitable Runge-Kutta method~\cite{shu_review}. 
The resulting scheme is strictly conservative, in the sense that the total mass, momentum and energy of discrete fields are strictly conserved,  exactly reproducing the properties of the NVE nonequilibrium (relaxation) trajectories.
Once density, momentum and energy are available, the other relevant fields (e.g. temperature, pressure, etc.) are evaluated though the EoS.}

\section{Results and discussion}
\label{sec:results_discussion}

For the analysis of the relaxation of the system with the initial conditions described in Sec.~ \ref{sec:computational_setup}, we focus on six field observables: number density, $\rho\left (\textbf{x},t\right )$ 
momentum, $\boldsymbol{\pi}\left (\textbf{x},t\right )$, temperature, $T\left (\textbf{x},t\right )$, energy, $e\left (\textbf{x},t\right )$, energy flux, $\boldsymbol{q}\left (\textbf{x},t\right )$, and pressure, $p \left (\textbf{x},t\right )$. 

\subsection{Atomistic results}
\label{sec:results_discussion-atomistic_results}
\begin{figure}[h!]
 \includegraphics[width=15cm]{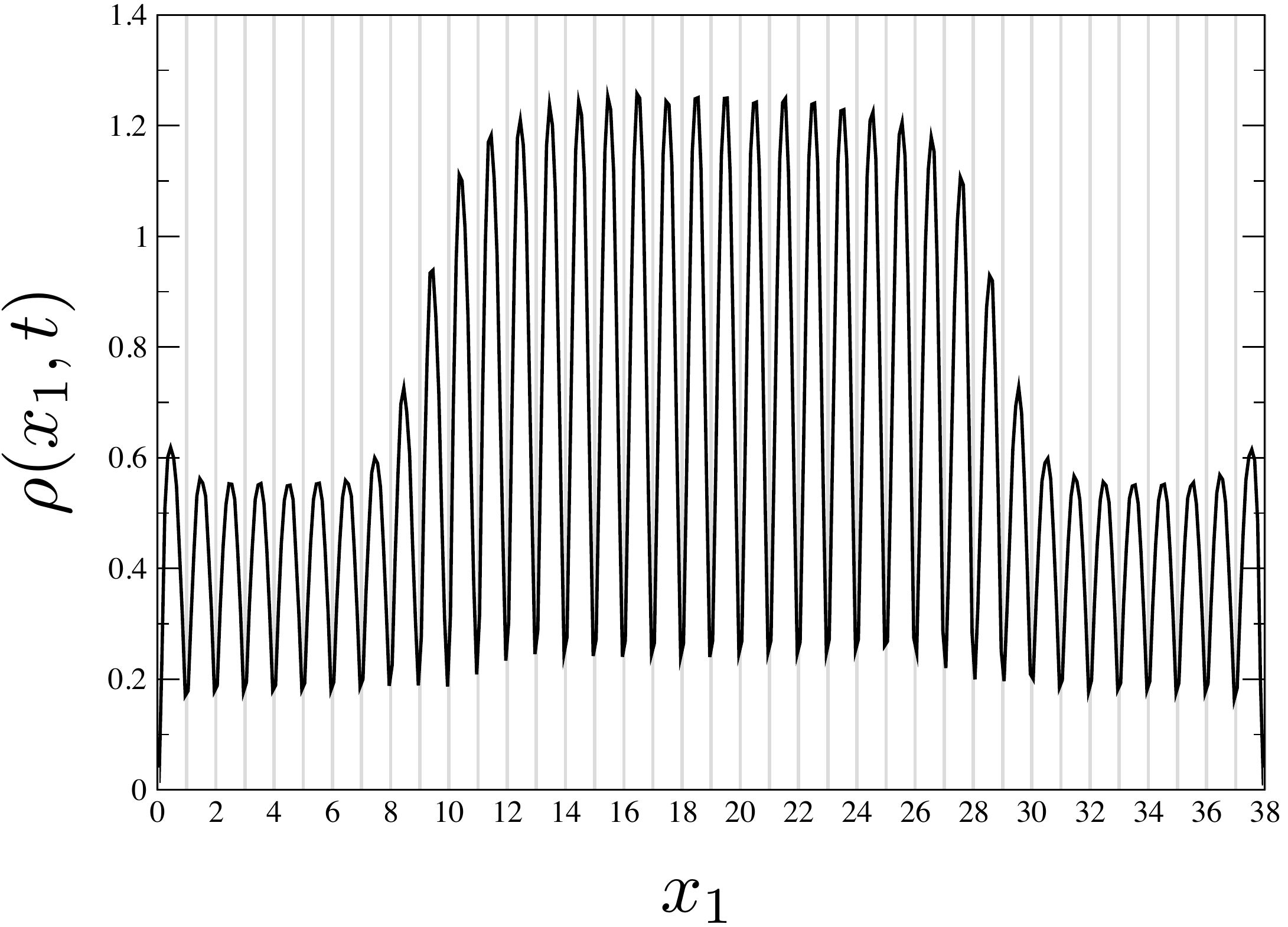}
 \caption{Initial density profile, $\rho\left ( \textbf{x}, 0\right )$, computed over the fine, $0.1~\sigma$ grid. The vertical grey lines represent the boundary of the original slice used to impose the initial condition.} 
 \label{fig:density_0-fineGrid}
\end{figure}

In the top panel of Fig.~\ref{fig:initial_rho} we report the density profile along $x_1$ at $t = 0$ computed on a (coarse) grid with a $\Delta x$ of $1~\sigma$. We report only the density profile along $x_1$ because, given the initial conditions, along $x_2$ and $x_3$ $\rho$, and all the other fields, are constant. At $t = 0$ the density shows the double $s$/mirror-s shaped profile explained in Sec.~\ref{sec:atomistic_setup}.
If we compute the density on a finer grid, with a step of $1/10~\sigma$, we notice that $\rho\left ({x}_1, 0\right )$  is not smooth (see Fig.~\ref{fig:density_0-fineGrid}). This is due to the effect of the soft confining walls discussed in Sec.~\ref{sec:atomistic_setup}, which produce a depletion at the slices boundaries, and a complementary increase at their center. $\rho$ can be made smoother in two different ways. First, we can use the non-mollified version of the density field to impose the initial condition, integrating the dynamics of the particles with integrators that can deal with impulsive forces~\cite{CiccottiImpulsiveIntegrator}. Second, we can sample the initial conditional PDF running several, independent RMDs with the mollified version of the density using grids shifted with respect to each other. In the limit of an infinite number of such grids $\rho\left ({x}_1,0\right )$ will be perfectly smooth. However, we expect that already with a small number of grids (say $5$ to $10$), shifted of a distance of the order of the range of the potential generated by the soft walls, $\rho\left ({x}_1,0\right )$ will be smooth. It is worth remarking that this approach is only apparently more expensive, because the length of each RMD would be $1/n$-th of the original one, where $n$ is the number of shifted grids. 
However, the objective of this work is not to describe the relaxation from a specific initial condition but rather to compare the macroscopic and microscopic representation of the relaxation of a system characterized by nonuniform, steep density profile. Thus, we have taken the simpler approach of initializing continuum simulations with atomistic fields on the fine grid (more details are given below).

\begin{figure}[h!]
 \includegraphics[width=15cm]{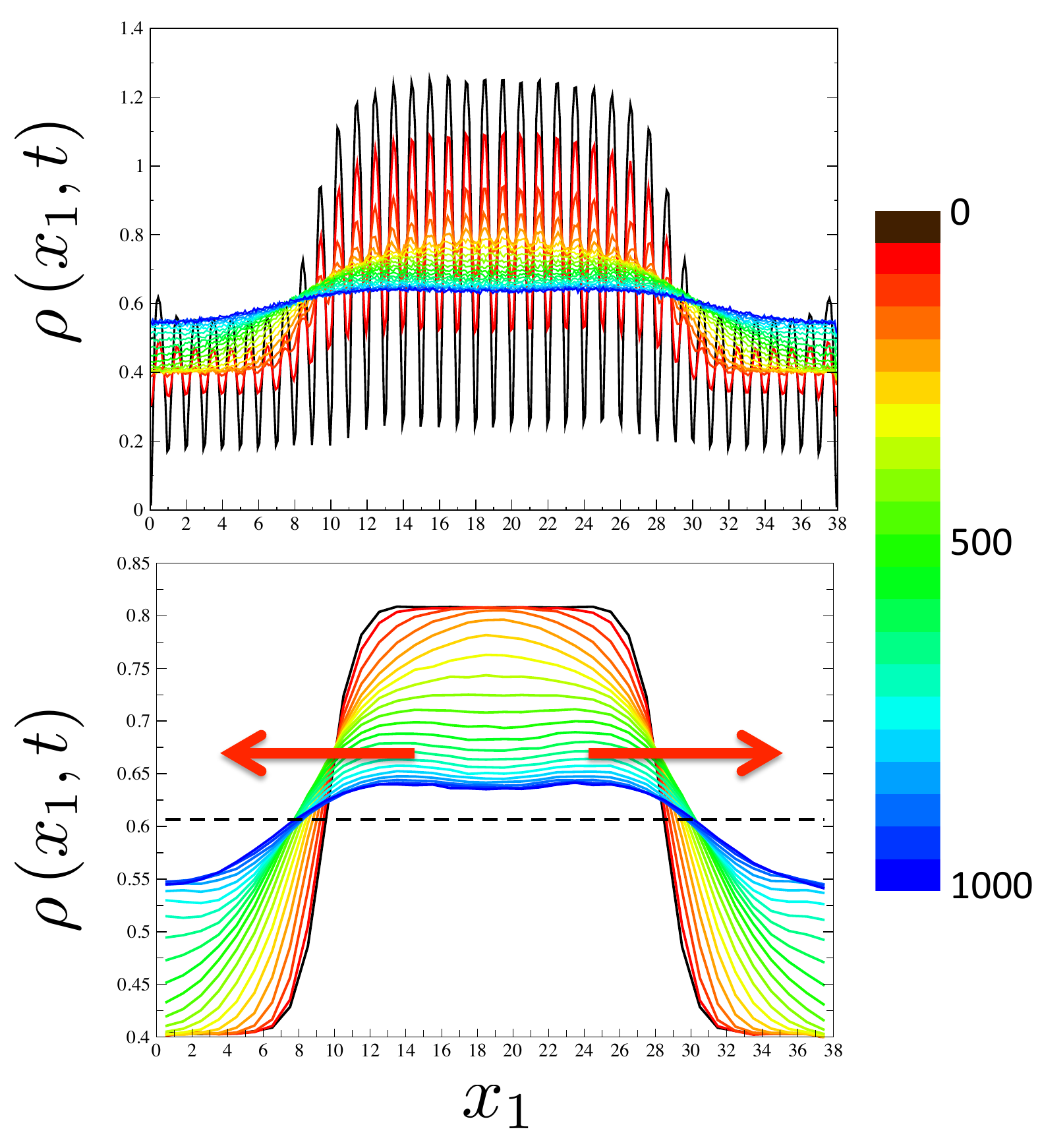}
 \caption{$\rho(x_1, t)$ for the first 1000 time steps computed on the fine (top) and coarse (bottom) grid. Line colors indicate the time $t$ of the corresponding density profile. At short $t$ lines are red, become blu at long $t$ passing by yellow and green. The dashed line denotes the value $\bar \rho = (\rho_1 + \rho_2)/2$. The low/high density domain interface is (arbitrary) defined as the (set of) point(s) $x^*_1$ at which $\rho(x^*_1, t) = \bar \rho$. The arrows indicate how $x^*_1$ changes with $t$.} 
 \label{fig:density_0-1000}
\end{figure}

Let us now move to the analysis of the atomistic mechanism of relaxation of the system.
In Fig.~\ref{fig:density_0-1000} we show $\rho\left (x_1,t\right )$ for $t \in \left [1,1000 \right]$. In the top panel we report the density on the fine grid, and in the bottom panel that on the coarse grid. We notice that the large density oscillations decay very quickly. In practice, within $\sim 200$ timesteps  the density field is smooth. As we will show below, these oscillations have effect only on the pressure field at short times. Thus, when not explicitly mentioned, we will discuss results computed on the coarse grid.

As a general remark, the dynamics of the density field follows the expected path: the (higher) density in the central region decreases, and the (lower) density in the peripheral regions  (connected by PBC) increases.  In $\sim 1000$ timesteps the density in the central region decreased of $\sim 40$~\% of the initial $\rho_1 - \rho_2$ (see Sec.~\ref{sec:atomistic_setup}). An analogous process of opposite sign occurs in the low density domain. 

\begin{figure}[h!]
 \includegraphics[width=15cm]{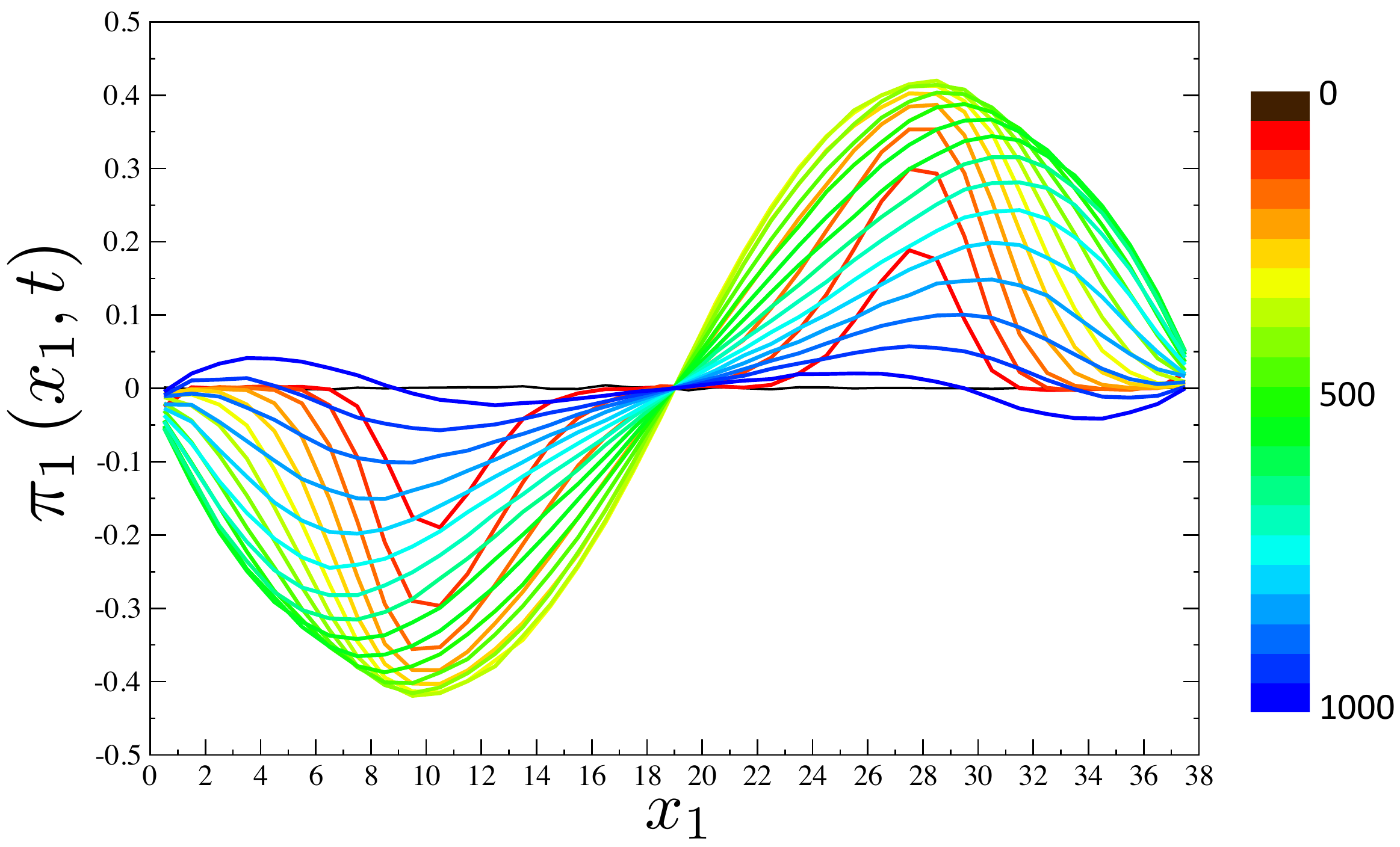}
 \caption{$\pi_1(x_1, t)$ for the first 1000 time steps. The color-coding is the same as in Fig.~\ref{fig:density_0-1000}} 
 \label{fig:momentum_0-1000}
\end{figure}

Analyzing more in detail the dynamics of $\rho\left (x_1,t\right )$ at short times ($t \leq 1000$), we notice that it follows two different regimes. Initially the density evolves forming a bell-like profile, with $\rho\left (x_1,t\right )$ higher at the center of the high density region and lower at the borders of the simulation box. Then, at $t \sim 300$,  the $\rho\left (x_1,t\right )$ starts to decrease faster at the center of the high density region than at the borders. At $t \sim 500$ the density profile in the central region is flat, and then becomes slightly concave. This trend can be explained analyzing the $x_1$ component of the momentum field, $\pi_1\left (x_1, t\right )$ (Fig.~\ref{fig:momentum_0-1000}). Obviously, at $t = 0$ the momentum is zero everywhere. At very short times ($t \sim 0 - 250$) the momentum is sizably different from zero only is correspondence of the ``interface'', i.e. in the region of significant gradient of the density field. In this time interval the sign of the momentum is negative at the left interface, and positive at the right interface. The combination of these characteristics of $\pi_1\left (x_1,t\right )$ produce the double effect of lowering the difference of the density between the center and periphery of the box, and moving the interface in the direction high $\rightarrow$ low density (see the arrows in Fig.~\ref{fig:density_0-1000}).

At longer times ($t \in \left [250, 900\right]$) the momentum fields takes non negligible values over all the $x_1$ domain. As before, $\pi_1\left (x_1,t\right )$ is negative in the left half box, and positive in the right one, and presents a ``node'' at $x_1 = 19$ (center of the box). The position of the maximum of the momentum field (in absolute value), $x_p^{max} =  \text{arg}\max \left ( |\pi_1\left (x_1,t\right )|\right )$, goes first in the direction center $\rightarrow$ periphery and then reverts. This induces a corresponding change in the (tiny) modes of the density field, $x_\rho^{max} = \text{arg} \max \ \left ( \rho \left (x_1,t\right )\right )$. 

\begin{figure}[h!]
\includegraphics[height=15cm]{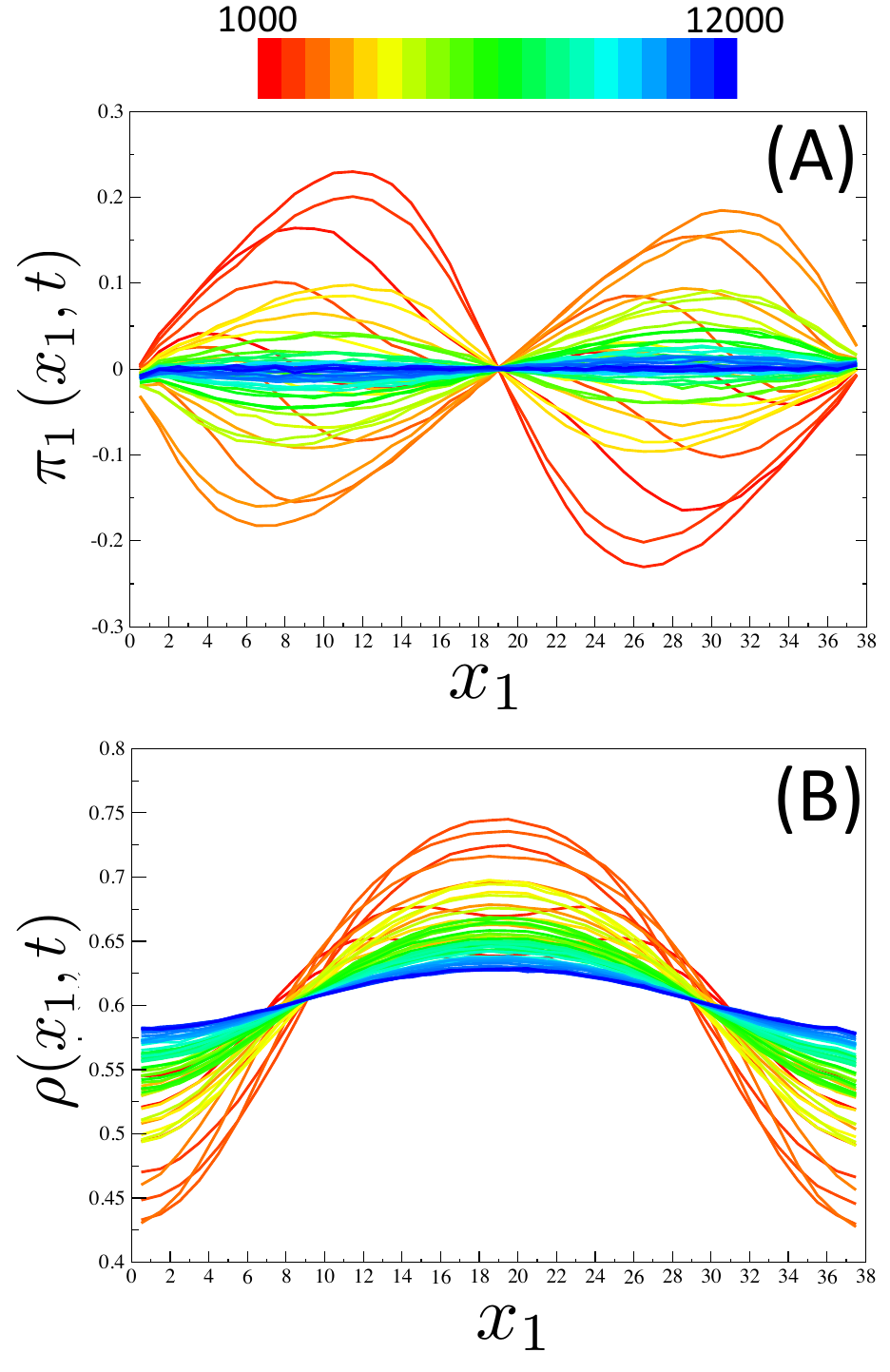}
 \caption{Momentum (A) and density (B) fields at longer times ($t \geq 1000$). Consistently with the other figures, the color of the of field profiles represent the time at which they are measured, from red ($1000$) to blue ($12000$).} 
 \label{fig:density_momentum_1000-12000}
\end{figure}

At even longer times, the momentum field first presents several nodes ($3$ in the last two curves of Fig.~\ref{fig:momentum_0-1000}), then revert sign with respect to $t = 0$ (see also Fig.~\ref{fig:density_momentum_1000-12000}/A), and evolves following a damped oscillating dynamics. At very long times $\pi_1\left (x_1,t\right )$ eventually converges to a uniform zero field. The origin of $\pi_1\left (x_1,t\right )$ sign inversion is the ``clash'' between the two interface fronts through PBCs. The $\pi_1\left (x_1,t\right )$ sign inversion, in turn, brings to an increase of $\rho \left (x_1,t\right )$ in the central part of the simulation box (see Fig.~\ref{fig:density_momentum_1000-12000}/B), thus increasing once again the density gradient. To this density gradient is associate a force which tends to restore the density uniformity. The alternation of these two phenomena produces the oscillatory behavior or the momentum field.

\begin{figure}[h!]
\includegraphics[height=15cm]{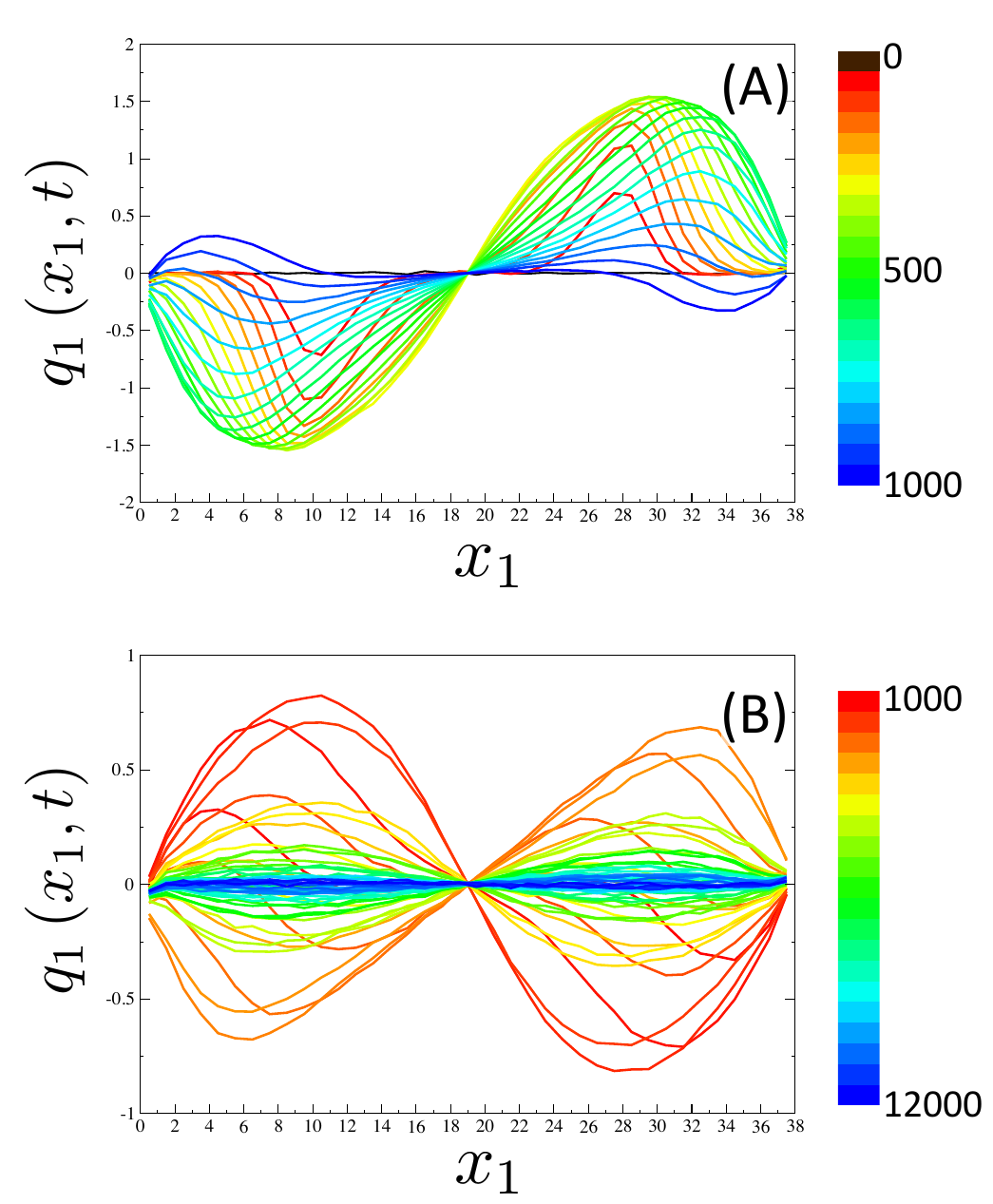}
 \caption{Energy flux field at (A) short ($t \leq1000$) and (B) long times.} 
 \label{fig:energyFlux-1_1000-1001_12000}
\end{figure}

As for the ``dynamics'' of the temperature field, initially the system is thermalized at the $T \sim 1.5$, and thus $T \left (x_1, 0\right )$ is uniform. However, as soon as the relaxation starts the temperature field becomes non uniform and, indeed, presents a complex profiles. Far from the interface, at the center of the high and low density regions, the temperature remains constant at the initial value. At the interface we observe two opposite behaviors: ahead of the interface the fluid gets warmer while behind it gets colder. To explain this observation we must first remark that at the beginning the relaxation process is very fast, thus we can consider that it is locally adiabatic, i.e. there is no (significant) exchange of heat within the fluid. This is confirmed by the profile of the energy flux, $j_1\left (x_1, t \right )$ (Fig.~\ref{fig:energyFlux-1_1000-1001_12000}/A), which is initially highly peaked at the interface. Thus, on the high density side of the interface the relaxation amounts to an adiabatic expansion (density is decreasing), and on the low density side to an adiabatic compression (density is increasing), which, according to classical thermodynamics, produces a decrease and increase of the temperature, respectively. The expansion (compression) keeps producing a decrease (increase) of the temperature on the high (low) density domain of the sample untill the temperature gradient becomes high enough  that a sizable inverse Fourier-like (i.e. thermal gradient driven) energy flux becomes effective in restoring a uniform $T \left (x_1, t\right )$ (Fig.~\ref{fig:energyFlux-1_1000-1001_12000}/B). 
 
 \subsection{Continuum calculations and comparison with atomistic simulations} 
 We considered two different initial conditions. One is that corresponding to the atomistic fields at $t = 0$ computed over the fine grid. In the following we shall refer to this case as ``rough initial conditions''. The other initial condition is that corresponding to atomistic fields at $t  = 300$, i.e. after the large oscillations of the atomistic fields have been ``absorbed'', hereafter named ``smooth initial conditions''.

\begin{figure}[h!]
%
%
\includegraphics[width=15cm]{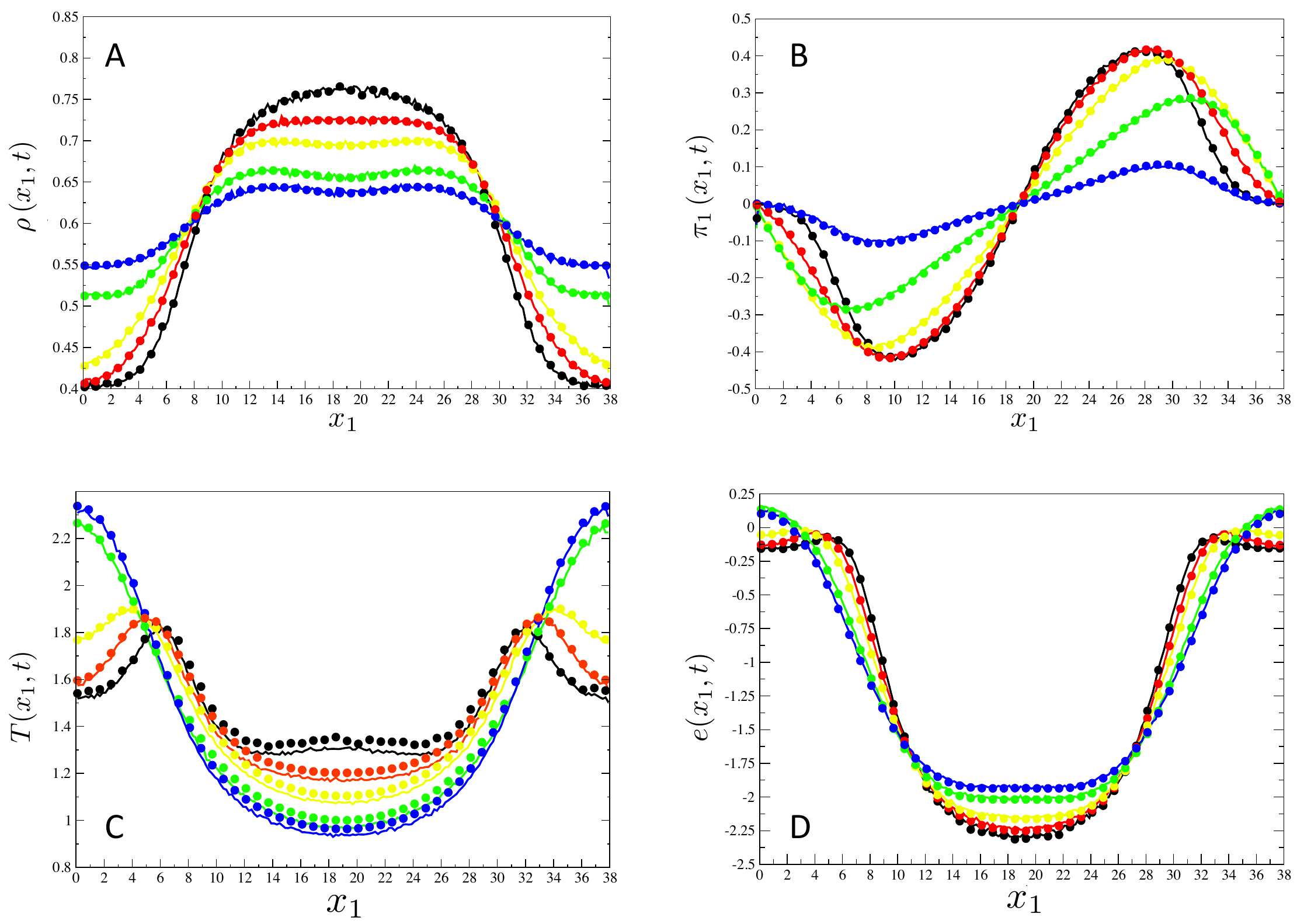}
 \caption{Comparison between selected atomistic (lines) and continuum (circles) fields. (A) $\rho(x_1, t)$; (B) $\pi_1(x_1, t)$; (C) $T(x_1, t)$; and (D) $E(x_1, t)$ (energy density) at short times  (corresponding to the time interval $300$-$900$ of Fig.~\ref{fig:density_0-1000}) starting from the smooth initial conditions. To make the small differences more visible, continuum data are plotted on a grid three times coarser than the one used in simulations.} 
 \label{fig:0-1000AC}
\end{figure}

Let us start our analysis from this latter case. In Fig.~\ref{fig:0-1000AC} we compare atomistic and continuum density, momentum and temperature fields. Atomistic and continuum density and momentum fields are in perfect agreement, while we note a mismatch in the temperature field. This is due to the accuracy of the EoS used in continuum simulations. In fact, as mentioned in the original article,~\cite{JohnsonZollwegGubbins} the accuracy of this EoS is lower at higher density, which explains why the difference between atomistic and continuum results is larger in the high density region and decreases with $t$, in parallel with the decrease of density in the central region. This comparison brings to the following conclusions. The matching between a selected list of atomistic and continuum fields is almost perfect. Some continuum fields present minor differences with atomistic one due to the limited accuracy of some of the empirical relations needed by continuum models. Nevertheless, these differences do not prevent to correctly describe interface phenomena involving length of the order of multiple interatomic distances ($\sim 10~\sigma$, the width of the interface, as measured by the length over which the density fields goes from the maximum to the minimum at the initial condition).

\begin{figure}[h!]
%
%
\includegraphics[width=15cm]{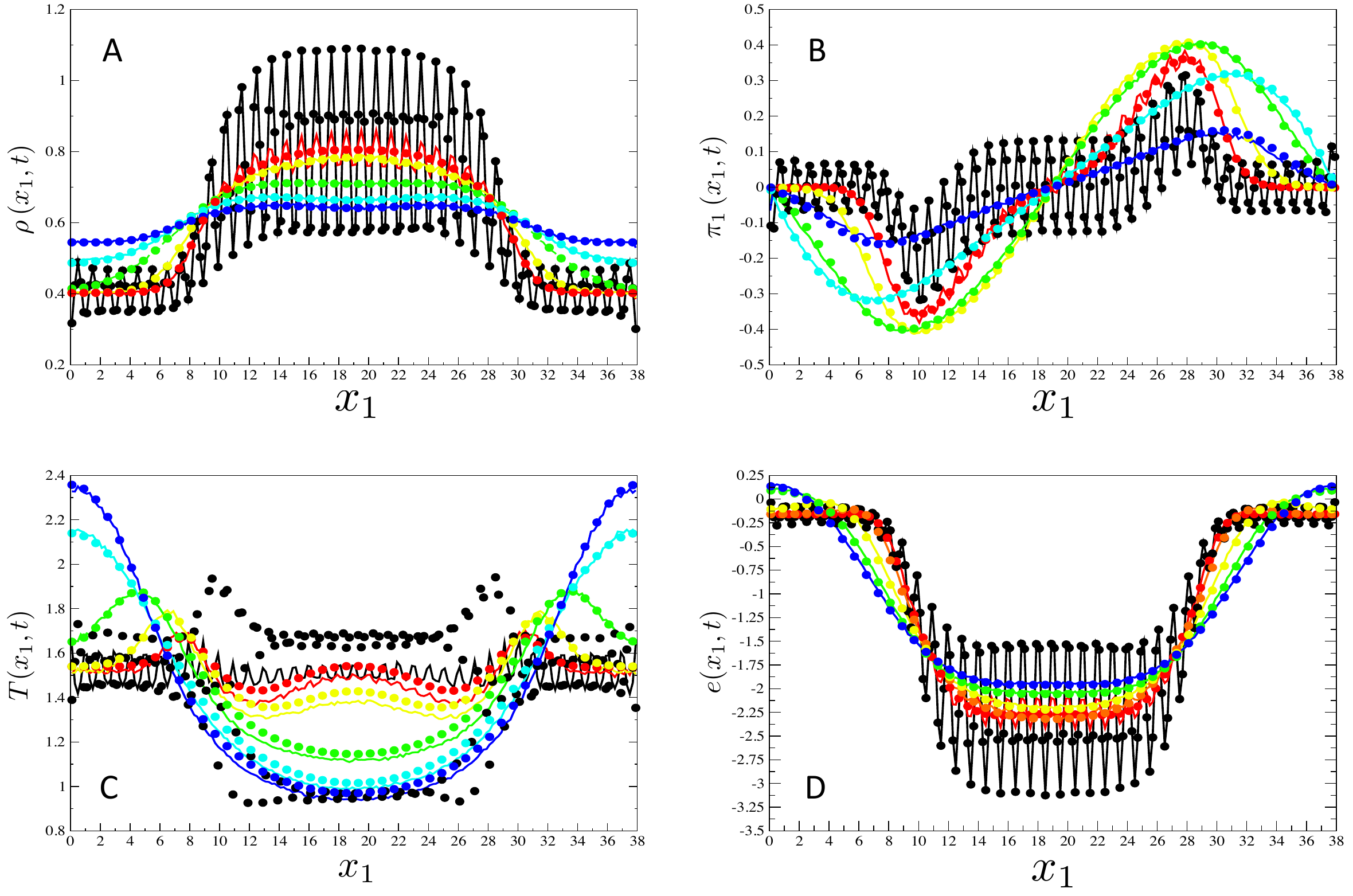}
 \caption{Same as in Fig.~\ref{fig:0-1000AC} but starting from the rough initial conditions. We remark the difference between the atomistic and continuum temperature in the initial condition. The black dots of panel (c) are only apparently scattered, while they denote very large oscillations of the continuum temperature field. In comparison, the corresponding atomistic system is characterized by   moderate oscillations. As explained in the main text, this is due to the accuracy of the EoS, which critically depend on the density of the system.} 
 \label{fig:0-1000AC-R}
\end{figure}

Let us continue by comparing  atomistic and continuum simulations when the initial conditions present characteristic lengths on the atomistic scales, i.e. when at $t = 0$ of the atomistic simulations the density presents oscillations of $\sim 1~\sigma$ wavelength (Fig.~\ref{fig:density_0-fineGrid}). Atomistic and continuum density, momentum and temperature fields at short times ($t \leq 1000$) are compared in Fig.~\ref{fig:0-1000AC-R}. We notice that atomistic and continuum $\rho(x_1, t)$ and  $\pi_1(x_1, t)$ fields match relatively well. The difference is visible at short times, where the atomistic field presents atomistic scale oscillations, while the continuum fields are smooth. This means that transport modeled by usual constitutive laws, with transport coefficients derived from bulk MD, is faster than transport at the nanoscale. The situation is very different for the temperature field. At the beginning, continuum $T(x_1, 0)$ presents very large oscillations, much larger than in the atomistic case. This is due to the limited accuracy of the EoS. Present results bring us to the conclusion that the continuum description of phenomena involving molecular lengths and time scales are qualitatively correct. However, there are fields, like the temperature field, that critically depend on the accuracy of empirical relations, which proven to fail also in the case of simple Lennard-Jones systems. This means that continuum theories might be inadequate to describe physical phenomena, such as ``collisions'' between fluids at very high energy, which might bring to large  fluctuations of fields on the molecular length scales.  

\section{Conclusions.}
\label{sec:conclusions}

In this work we applied dynamical nonequilibrium molecular dynamics to study the relaxation process of a fluid in presence of a large density gradient. We compared atomistic results against fields obtained from continuum theories. This example shows that phenomena involving lengthscales of the order of ten times the molecular scale are well described by continuum theories. When even shorter lengthscales are involved, the continuum theories can partly fail due to the limited accuracy of some of the empirical relations used in the macroscopic models.


\section*{Acknowledgements}
GC and SM acknowledge financial support from 
the Istituto Italiano di Tecnologia under the SEED project grant No. 259 SIMBEDD -- Advanced 
Computational Methods for Biophysics, Drug Design and Energy Research. 
S.M. acknowledges financial support from the MIUR-FIRB Grant No. RBFR10ZUUK.
The research leading to these results has received funding from the European Research Council under the European Union's Seventh Framework Programme (FP7/2007-2013)/ERC Grant agreement n$^\circ$ [339446].
The authors thank the ICHEC and the CINECA Supercomputing Centres for the provision of computational resources.


%

\newpage

\begin{figure}[h!]
\includegraphics[height=15cm]{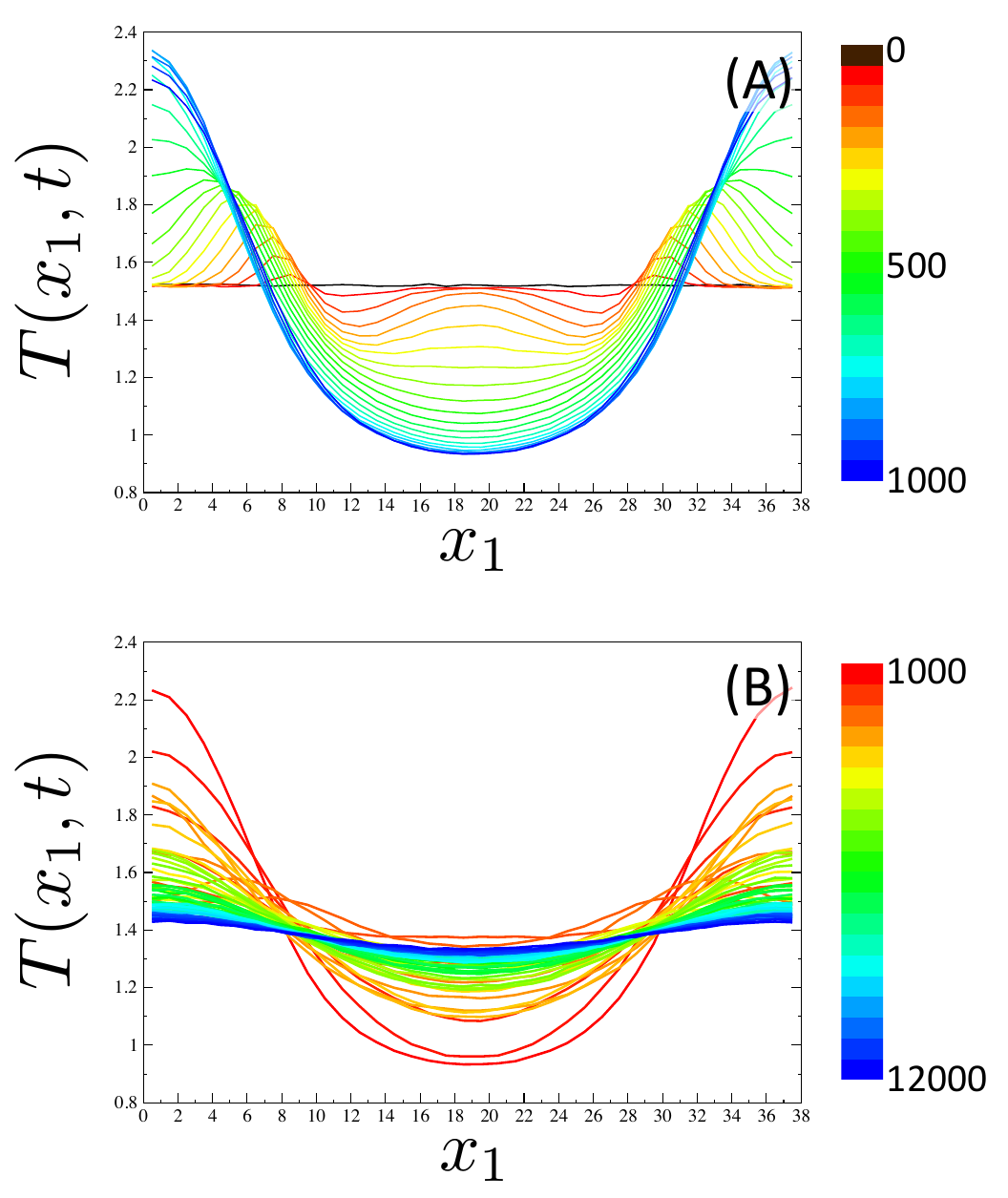}
 \caption{Temperature field at (A) short ($t \leq1000$) and (B) long times.} 
 \label{fig:temperature-1_1000-1001_12000}
\end{figure}

\end{document}